\documentclass[prl,aps,twocolumn,showpacs,preprintnumbers,amsmath,amssymb]{revtex4}

\usepackage{mathrsfs}
\usepackage{amsmath}
\usepackage{amssymb}
\usepackage{graphicx}
\usepackage{color}
\usepackage{dcolumn}
\usepackage{bm}

\makeatletter
\newcommand{\rmnum}[1]{\romannumeral #1}
\newcommand{\Rmnum}[1]{\expandafter\@slowromancap\romannumeral #1@}

\makeatother

\begin{document}

\title{Evidence for a heavy-mass surface state in SmB$_6$: Magneto-thermoelectric transport on the (011)-plane}
\author{Yongkang Luo$^{1}$\footnote[1]{Electronic address: ykluo@lanl.gov}, Hua Chen$^{2,3}$, Jianhui Dai$^{4}$, Zhu-an Xu$^{5}$, and J. D. Thompson$^{1}$}
\address{$^1$Los Alamos National Laboratory, Los Alamos, New Mexico 87545, USA;}
\address{$^2$International Center for Quantum Materials and School of Physics, Peking University, Beijing 100871, China;}
\address{$^3$Collaborative Innovation Center of Quantum Matter, Beijing 100871, China;}
\address{$^4$Department of Physics, Hangzhou Normal University, Hangzhou 310036, China;}
\address{$^5$Department of Physics, Zhejiang University, Hangzhou 310027, China.}

\date{\today}

\begin{abstract}

Motivated by the high sensitivity to Fermi surface topology and scattering mechanisms in magneto-thermoelectric transport, we have measured the thermopower and Nernst effect on the (011)-plane of the proposed topological Kondo insulator SmB$_6$. These experiments, together with electrical resistivity and Hall effect measurements, suggest that the (011)-plane also harbors a metallic surface with an effective mass on the order of 10-10$^2$ $m_0$. The surface and bulk conductances are well distinguished in these measurements and are categorized into metallic and non-degenerate semiconducting regimes, respectively. Electronic correlations  play an important role in enhancing scattering and also contribute to the heavy surface state.

\end{abstract}


\maketitle

\section{Introduction}

Topological insulators (TIs) represent a new form of quantum matter: in a single material, the sample's bulk is insulating with an energy gap that is traversed by an intrinsically metallic Dirac surface state in which the electron spin is locked perpendicular to the crystal momentum by  strong spin-orbit coupling (SOC)\cite{Fu-PRL07,Moore-07,Qi-PRB08,Qi-RMP}. Such a non-trivial surface state, protected by time-reversal symmetry, has potential for applications such as spintronics and Majorana fermions on the interface between a TI and a superconductor\cite{Moore-TI}. Recently, the intermediate-valence compound SmB$_6$ has been proposed both theoretically and experimentally to be a topological Kondo insulator (TKI)\cite{Dzero-PRL2010,Takimoto-SmB6,Dzero-PRB2012,Alexandrov-PRL2013,Wolgast-RT,Botimer-Hall,Lu-SmB6,Kim-NM,Xu-Spintex}. In contrast to conventional TIs, the topologically non-trivial state in a TKI stems from electronic correlations. Specifically, Kondo hybridization between a narrow $f$-band and a broad $d$-band produces a bulk energy gap at low temperatures, and the Fermi level resides in this gap\cite{Takimoto-SmB6,Lu-SmB6,Feng-TKI,Nikolic-HFTKI}. By necessity, hybridization between $f$- and $d$-electrons is an odd function of momentum to preserve time-reversal symmetry, and consequently, the hybridization matrix elements vanish at high symmetry points in the Brillouin zone, leading to three Dirac points on the surface.

SmB$_6$ crystallizes in a CsCl-type crystalline phase in which a B$_6$-octahedral cluster and Sm form a simple cubic structure.
Non-local resistance\cite{Wolgast-RT} and surface Hall effect\cite{Botimer-Hall} measurements on the (001)-plane of SmB$_6$ revealed the existence of a metallic surface conduction in parallel with an insulating bulk, which reconciles the long-standing mystery of a saturating low-temperature resistivity. Though these experimental techniques have been quite useful, and can be performed at temperatures well below the lowest temperature capabilities of angle-resolved photoemission spectroscopy (ARPES)\cite{Xu-SmB6ARPES,Hasan-SmB6ARPES,Xu-Spintex,Xu-SmB6ARPES2}, they are unable to to provide details of the metallic surface states. More information is contained in measurements of magneto-thermoelectric transport.

In the presence of a temperature gradient $-\nabla T$, an electric field $\textbf{E}$ and a magnetic field $\textbf{B}$, the total current density $\textbf{J}$=$\boldsymbol{\sigma}$$\cdot$$\textbf{E}$+$\boldsymbol{\alpha}$$\cdot$$(-\nabla T)$, where $\boldsymbol{\sigma}$ is the conductivity tensor, and $\boldsymbol{\alpha}$=$AT\frac{\partial\boldsymbol{\sigma}}{\partial\varepsilon}|_{\varepsilon=\varepsilon_F}$ is the Peltier conductivity tensor ($A$=$\frac{\pi^2k_B^2}{3q}$, $k_B$ is Boltzmann's constant, $q$ is the charge of carriers)\cite{Ziman}. In an equilibrium state without net current flow, the Boltzmann-Mott transport equation yields the thermoelectric tensor
\begin{equation}
\textbf{S}=\boldsymbol{\alpha}\cdot\boldsymbol{\sigma}^{-1}=AT \frac{\partial\ln\boldsymbol{\sigma}}{\partial\varepsilon}|_{\varepsilon=\varepsilon_F}.
\label{Eq.1}
\end{equation}
Expanding Eq.~\ref{Eq.1} gives the thermopower (diagonal) and Nernst (off-diagonal) signals\cite{Ong-PbSnSe}, i.e.,
\begin{subequations}
\begin{align}
S_{xx}(B)&=AT(\frac{\sigma_{xx}^2}{\sigma_{xx}^2+\sigma_{xy}^2}D_{xx}+\frac{\sigma_{xy}^2}{\sigma_{xx}^2+\sigma_{xy}^2}D_{xy}), \label{Eq.2a}\\
S_{xy}(B)&=AT\frac{\sigma_{xx}\sigma_{xy}}{\sigma_{xx}^2+\sigma_{xy}^2}(D_{xy}-D_{xx}),\label{Eq.2b}
\end{align}
\end{subequations}
in which $D_{ij}$=$\partial\ln\sigma_{ij}/\partial\varepsilon|_{\varepsilon=\varepsilon_F}$. Being the energy derivative of $\sigma_{ij}$, $S_{ij}$ is more sensitive to Fermi surface topology and scattering mechanisms (viz. $\tau$$\propto$$\varepsilon^{\lambda}$, $\tau$ is relaxation time) than $\sigma_{ij}$ itself\cite{Ong-PbSnSe,Zhu-Graphite,Fauque-Bi2Se3,Delves-Thermomag}.

By determining elements of the conductivity and thermoelectric tensors, we find that the (011)-plane of SmB$_6$ also harbors a metallic surface state at low temperatures\cite{Wolgast-Corbino}, but there are no clear signatures for quantum oscillations, in contrast to reported oscillations from torque magnetometry measurements\cite{Li-SmB6Torque}. Thermoelectric transport reveals non-degenerate semiconducting behavior above 10 K, consistent with the presence of a bulk Kondo hybridization gap. The magneto-thermoelectric tensor exposes scattering mechanisms of the bulk and surface and, significantly, evidence for a heavy effective mass of the surface state that was predicted theoretically\cite{Alexandrov-PRL2013}.

\section{Experimental details}

High-quality single crystals of SmB$_6$ were synthesized by an aluminium-flux method with the starting materials: samarium ingot ($\geq$99.99\%, Metall), boron powder ($\geq$99.999\%, Alfa Aesar) and aluminium granules ($\geq$99.99\%, Alfa Aesar). Sm, B and Al were weighted in an atomic ratio of 1:4:200. The mixture was placed in an alumina crucible and heated to 1500 \textordmasculine C in a flowing-helium atmosphere. During the reaction, the crucible was covered with an alumina lid to reduce Al evaporation. After maintaining 1500 \textordmasculine C for 2 hours, the furnace was slowly cooled to 620 \textordmasculine C in two weeks. The aluminium flux was dissolved by a concentrated NaOH solution in a fume hood, and shiny single crystals of SmB$_6$ of millimeter-size were picked out.

\begin{figure}[htbp]
\includegraphics[width=8.0cm]{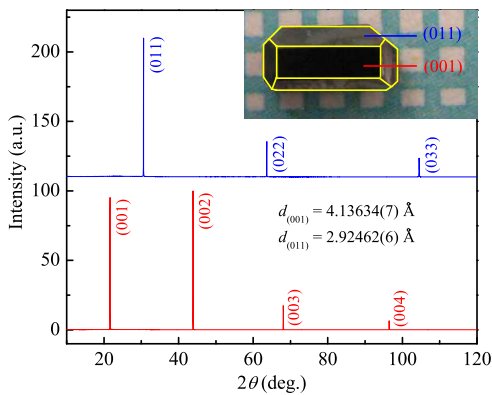}
\caption{(Color online)\label{Fig.1} XRD characterization of the SmB$_6$ single crystal on ($001$)- and ($011$)-orientations. The inset shows a picture of an as-grown SmB$_6$ single crystal with both ($001$)- and ($011$)-planes.}
\end{figure}

The sample quality was checked by X-ray diffraction (XRD, Cu-$K_{\alpha 1}$) as shown in Fig.~\ref{Fig.1}. Only ($00l$) ($l$=1,2...) peaks can be observed on the XRD pattern (red curve), while no impurity phases (e.g. Al) can be detected. Some as-grown samples have both natural (001)- and (011)-crystalline surfaces as depicted in the inset to Fig.~\ref{Fig.1}. We carefully polished the single crystals to the (011)-orientation using Al$_2$O$_3$ sand-papers with increasingly finer grits. The final step of polishing was performed with a high-speed mini-grinder with superfine diamond paste (0.5 $\mu$m) to get a mirror-like shiny surface. The orientation of the polished sample was confirmed by XRD (blue curve), in which only ($0ll$) peaks are seen. The ratio $d_{(001)}/d_{(011)}$=1.414317..., indicating the precise (011)-orientation.

Ohmic contacts were made in a Hall-bar geometry, and both electrical resistivity and Hall resistivity were measured by an LR-700 AC resistance bridge. Thermoelectric transport measurements were carried out by means of a steady-state technique. For temperatures above 2 K, a pair of well calibrated differential Chromel-Au$_{99.93\%}$Fe$_{0.07\%}$ thermocouples was used to measure the temperature gradient; whereas, for temperatures below 2 K, the measurement was done by the standard one-heater-two-thermometer technique. Upon a thermal gradient $-\nabla T$$\parallel$$\textbf{x}$ and a magnetic field $\textbf{B}$$\parallel$$\textbf{z}$, both thermopower signal $S_{xx}$=$-E_x/|\nabla T|$ and Nernst signal $S_{xy}$=$E_y/|\nabla T|$ were collected by scanning field at fixed temperatures. All these measurements were made on the same crystal.

\section{Results and Discussion}

\begin{figure*}[htbp]
\includegraphics[width=16cm]{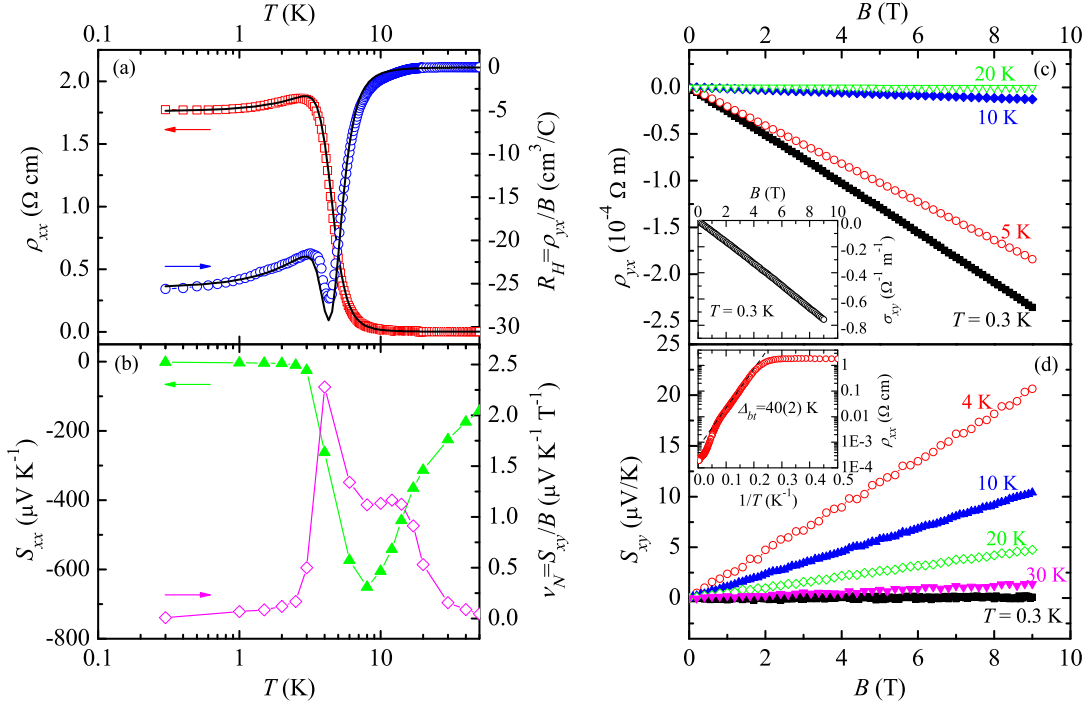}
\caption{(Color online)\label{Fig.2} Electrical and thermoelectric transport properties of SmB$_6$ on the (011)-plane, plotted versus temperature. (a) The resistivity ($\rho_{xx}$) and Hall coefficient ($R_H$=$\rho_{yx}/B$, $B$=9 T). (b) The thermopower ($S_{xx}$) and Nernst coefficient ($\nu_N$=$S_{xy}/B$, $B$=9 T). The solid lines in (a) are fits to the two-channel model discussed in the text. (c) and (d) respectively shows the field dependence of $\rho_{yx}$ and $S_{xy}$ at selected temperatures. The inset to (c) displays the Hall conductivity $\sigma_{xy}(B)$ at 0.3 K; while the inset to (d) is an Arrhenius plot of $\rho_{xx}(T)$, which shows thermal activation across a transport gap $\Delta_{bt}$=40(2) K. }
\end{figure*}

The temperature dependencies of resistivity and the Hall coefficient are shown in Fig.~\ref{Fig.2}(a). Similar to  previous measurements on the ($001$)-plane\cite{Wolgast-RT,Botimer-Hall,Syers-SmB6Gate}, metallicity is poor at room temperature. Upon cooling, $\rho_{xx}(T)$ rises with a thermally activated dependence but starts to level off below 4 K. The residual resistivity ratio $RRR$$\equiv$$\rho_{xx}$(0.3 K)/$\rho_{xx}$(300 K) is approximately 10$^4$ [see the inset to Fig.~\ref{Fig.2}(d)]. Significantly, $\rho_{xx}(T)$ displays a positive slope below 3 K, demonstrating an intrinsic metallic  conduction at low temperatures. Correspondingly, the Hall coefficient $R_H$ ($R_H$$<$0) below 3 K is also slightly temperature dependent. We emphasize that the Hall resistivity $\rho_{yx}(B)$ for $T$$<$3 K is almost strictly linear with magnetic field up to 9 T [see Fig.~\ref{Fig.2}(c)], indicative of effectively single-band conduction with low mobility ($\mu_s B$$\ll$1, see below). In this situation, the temperature dependence of the mobility for metallic conduction should be taken into account. For additivity in a parallel-resistor-like analysis of these data, we convert the resistivity tensor to a conductivity tensor, viz. $\boldsymbol{\sigma}$=$\boldsymbol{\rho}^{-1}$, where $\boldsymbol{\sigma}$ is a sum of both surface and bulk contributions, i.e.,
\begin{equation}
\sigma_{ij}=\frac{\rho_{ji}}{\rho_{xx}^2+\rho_{yx}^2}=\sigma^s_{ij}/t+\sigma^b_{ij},
\label{Eq.3}
\end{equation}
and $t$=68 $\mu$m is the thickness of the polished sample. From the Drude model, the diagonal and off-diagonal elements of $\boldsymbol{\sigma}$ are:
\begin{subequations}
\begin{align}
\sigma_{xx}&=\frac{2n_s|e|}{t}\frac{\mu_s(T)}{1+\mu_s^2(T)B^2}+n_b(T)|e|\frac{\mu_b}{1+\mu_b^2B^2},\label{Eq.4a}\\
\sigma_{xy}&=\frac{2n_se}{t}\frac{\mu_s^2(T)B}{1+\mu_s^2(T)B^2}+n_b(T)e\frac{\mu_b^2B}{1+\mu_b^2B^2},\label{Eq.4b}
\end{align}
\end{subequations}
in which $e$ is the charge of an electron, $n$ and $\mu$ are carrier density and mobility, respectively, and the subscript {\it s} (or {\it b}) denotes the surface (or bulk) contribution. According to Matthiessen's rule\cite{Ashcroft-SSP}, the temperature dependence of $\mu_s$ is treated as $\frac{1}{\mu_s(T)}$=$\frac{1}{\mu_{s0}}(1+c T^{\gamma})$. Whereas, we neglect a possible temperature dependence of $\mu_b$ for the bulk conduction because the thermal activation of $n_b(T)$=$n_{b0} \exp(-\Delta_{bt}/T)$ dominates. $n_s$, $\mu_{s0}$, $c$, $\gamma$, $n_{b0}$, $\Delta_{bt}$ and $\mu_b$ are free parameters in a fit to this  two-channel model of electrical transport. Results of the fit are converted back to a resistivity tensor and plotted as solid curves in Fig.~\ref{Fig.2}(a). Best fit parameters are listed in Table~\ref{Tab.1}. The obtained mobilities of both bulk and surface states are low.

\begin{table*}
\tabcolsep 0pt \caption{\label{Tab.1} Physical parameters for surface and bulk properties of SmB$_6$, compared with Bi$_2$Te$_2$Se\cite{Ren-Bi2Te2Se,Ong-Bi2Te2Se1}. }
\vspace*{-12pt}
\begin{center}
\def\temptablewidth{2.0\columnwidth}
{\rule{\temptablewidth}{1pt}}
\begin{tabular}{cccc|c||cc|c}
\textbf{Surface}~~~~~~     & SmB$_6$      &               &         &~Bi$_2$Te$_2$Se~~&~\textbf{Bulk}~~  & SmB$_6$~    &~Bi$_2$Te$_2$Se~\\ \hline
$n_s$~~~~~~~~~~~~~~  &8.35(4)$\times10^{14}$ cm$^{-2}$&  &   &~1.8$\times10^{12}$ cm$^{-2}$~~&$n_{b0}$~~~~&4.50(3)$\times10^{20}$ cm$^{-3}$~&~2.6$\times10^{16}$ cm$^{-3}$~\\
$\mu_{s0}$~~~~~~~~~~~~~~ &14.5(4) cm$^2$/Vs &          &           &~3200 cm$^2$/Vs~~&~$\mu_b$~~~~~~    &46.3(5) cm$^2$/Vs~            &~50 cm$^2$/Vs~\\
$c$~~~~~~~~~~~~~~~~                           &0.015(2) &         &           &~-~~&~$\Delta_{bt}$~~~~~~&40(2) K~                   &~580 K~\\
$\gamma$~~~~~~~~~~~~~~~~                      &1.5(2) &         &            &~-~~&~$\Delta_b$~~~~~~ &14 meV\cite{Hasan-SmB6ARPES}~&~350 meV~\\
$\lambda_s$~~~~~~~~~~~~~~~                    &$-$0.96(3)& $-$0.96(3)~~~  & $-$0.96(3)$^{\dag}~$    &~-~~&~$\lambda_b$~~~~~~ &$-$1.7(3)~                      &~-~\\
$\varepsilon_F$~~~~~~~~~~~~~~                 &0.46(2) meV & 0.46(2) meV~~~   &0.46(2) meV~            &~186 meV~~&                     &                           &\\
$k_F$=$\sqrt{4\pi n_s}$~~~                    &1.02 \AA$^{-1}$&0.1 \AA$^{-1}$\cite{Xu-SmB6ARPES}~~~&0.1 \AA$^{-1}$\cite{Xu-SmB6ARPES}~ &~0.047~~\AA$^{-1}$&       &     &\\
$v_F$=$\frac{\varepsilon_F}{\hbar k_F}$~~~~~~~&68 m/s       &700 m/s~~~  &1400 m/s~           &~6$\times10^{5}$ m/s~~&                   &     &\\
$l$=$\frac{\hbar k_F\mu_{s0}}{|e|}$~~~~~      &9.8 nm       &100 nm~~~   &200 nm~               &~79 nm~~&                   &   &\\
$m^*$=$\frac{\hbar k_F}{v_F}$~~~~~~           &17400 $m_0$  &165 $m_0$~~~ &83 $m_0$~       &~0.089 $m_0$~~&             &  &\\
$k_Fl$~~~~~~~~~~~~~                           &100          &100~~~      &200~               &~41~~&                   &   & \\
\end{tabular}
{\rule{\temptablewidth}{1pt}}
\end{center}
\vspace*{-5pt}
~~$^\dag$ In this column,  quantities are calculated in the case of a 2D quadratic dispersion, $N(\varepsilon)$$\propto$$\varepsilon^0$, $v^2$$\propto$$\varepsilon$, and thus Eq.~\ref{Eq.8} is unchanged, which leads to the same $\varepsilon_F$. Note that now $m^*$=$\frac{\hbar^2k_F^2}{2\varepsilon_F}$, and $v_F$=$\hbar k_F/m^*$.~~~~~~~~~~~~~~~~~~~~~~~~~~~~~~~~~~~~~~~~~~~~~~~~~\\
\end{table*}

\begin{figure}[htbp]
\includegraphics[width=8.0cm]{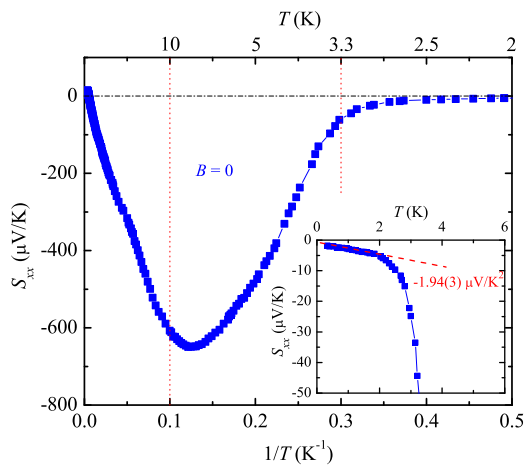}
\caption{(Color online)\label{Fig.3} Temperature dependence of the thermopower $S_{xx}$ plotted against inverse temperature $1/T$. The inset displays the low temperature variation of $S_{xx}(T)$. }
\end{figure}

Fig.~\ref{Fig.3} shows the temperature dependence of $S_{xx}$ measured at $B$=0. On the whole, $S_{xx}(T)$ is negative for $T$$<$195 K, in agreement with electron-type conduction and the sign of $R_H$. For $T$$>$195 K, $S_{xx}(T)$ changes sign and tends to saturate. At 300 K, the magnitude of $S_{xx}$ is about 15 $\mu$V/K, a typical value for a poor metal. Below 195 K, there are three distinct regimes for $S_{xx}(T)$. (\rmnum{1}) For $T$$>$10 K, $S_{xx}(T)$ is approximately linear with inverse temperature, which is characteristic of thermal activation of carriers across a bulk gap in an intrinsic semiconductor. The emergence of a Kondo hybridization gap in SmB$_6$ below 100 K has been reported in ARPES\cite{Xu-SmB6ARPES, Hasan-SmB6ARPES,Xu-SmB6ARPES2}, point-contact spectroscopy\cite{Zhang-SmB6PCS} and scanning tunneling spectroscopy\cite{Ruan-SmB6STM}. (\rmnum{2}) With decreasing $T$, $S_{xx}(T)$ develops a pronounced negative peak near 8 K, where $S_{xx}$ reaches $-$650 $\mu$V/K. (\rmnum{3}) With further cooling, the magnitude of $S_{xx}$ drops rapidly and starts to flatten below $\sim$3.3 K. Similar results were also seen in earlier measurements for $T$$>$2 K\cite{Sluchanko-SmB6S}. We show an enlarged view of the low temperature region in the inset to Fig.~\ref{Fig.3}. Clearly, $S_{xx}(T)$ approaches the origin linearly in the zero-temperature limit. The large magnitude of the slope $S_{xx}/T|_{T\rightarrow 0}$=$-1.94(3)~\mu$V/K$^2$ implies a very small Fermi energy $\varepsilon_F$, which will be discussed later. The pronounced negative peak at 8 K may be reminiscent of phonon drag. However, the Debye temperature $\Theta_D$ of SmB$_6$ is $\sim$400 K (estimated from LaB$_6$ in Ref.~\cite{Tanaka-LaB6}), and phonon drag is expected to be influential down to $\sim$40 K ($\Theta_D/10$) which is much higher than 8 K. Moreover, the rate of increase in $|S_{xx}(T)|$ in the region 3 K$<$$T$$<$8 K is far beyond the $(T/\Theta_D)^3$-law for typical phonon drag effects. An anomalous phonon drag effect, namely Umklapp processes\cite{Ashcroft-SSP}, should also be ruled out, since here the sign of this peak is consistent with the carrier type. With the concurrence of an approach to saturation below 3 K in $\rho_{xx}(T)$ and $S_{xx}(T)$, the rapid change of $|S_{xx}(T)|$ at 3 K$<$$T$$<$8 K signals the crossover from bulk-dominated conduction at $T$$\geq$8 K to surface-dominated conduction at $T$$\leq$3 K.

We now turn to the Nernst effect. For 0.3 K$\leq$$T$$\leq$100 K, the Nernst signal $S_{xy}(B)$ is nearly linear with $B$ [see Fig.~\ref{Fig.2}(d)], providing further evidence for the low carrier mobility of both surface and bulk states\cite{Ong-PbSnSe}. Within experimental resolution, there is no clear evidence of quantum oscillations in either $S_{xx}(B)$ or $S_{xy}(B)$ up to $B$=9 T. Previous magnetoresistivity measurements at temperatures to 40 mK and fields to 50 T also failed to resolve the Shubnikov-de Hass oscillations\cite{Li-SmB6Torque,Kebede-SmB650T}. The absence of quantum oscillations may not be surprising. On one hand, the mobility of the surface state is low so that a very high field would be required to quantize Landau levels; on the other hand, a magnetic field is detrimental to the Kondo effect, so in-gap surface states die away concomitantly with increasing field. The temperature dependence of the Nernst coefficient, defined as $\nu_{N}$=$S_{xy}/B$, is shown in Fig.~\ref{Fig.2}(b). $\nu_N$ is positive for the whole temperature range (note that we have taken the conventional definition\cite{Bridgman-Nernst} for the sign of Nernst  coefficient). $\nu_N(T)$ is negligibly small for $T$$>$100 K, but rises rapidly below 50 K and forms a plateau near 12 K with a value of about 1.17 $\mu$V/(K$\cdot$T). At 4 K, a sharp peak develops on top of this plateau and then $\nu_N(T)$ drops quickly to a small value below 3 K. Indeed, for a single-band, non-superconducting and non-magnetic metal, the Nernst coefficient is vanishingly small due to the so-called Sondheimer cancellation\cite{Sondheimer}. There are a few exceptions, however, where $\nu_N$ can be large\cite{Bel-NbSe2Nernst,Xu-Nernst,Wang-Nernst,Lee-CuCr2Se4,Sun-CeCu2Si2}, including the ambipolar effect in multi-band systems\cite{Bel-NbSe2Nernst} and the asymmetry of on-site Kondo scattering in Kondo-lattice systems\cite{Sun-CeCu2Si2}.


\begin{figure}[htbp]
\includegraphics[width=8cm]{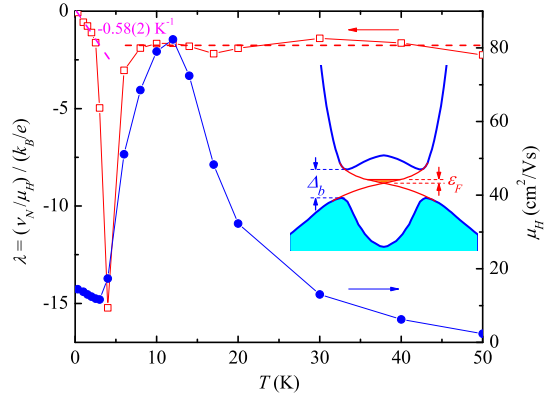}
\caption{(Color online)\label{Fig.4} The dimensionless ratio $\lambda$=$(\nu_N/\mu_H)/(k_B/e)$ as a function of $T$. Also shown is the Hall mobility $\mu_H$. The inset is a schematic sketch of the band structure.}
\end{figure}

In the low-field limit ($\mu B$$\ll$1, $\sigma_{xy}$$\ll$$\sigma_{xx}$), Eq.~\ref{Eq.2b} reduces to
\begin{equation}
\nu_N(T)=AT\mu_H\frac{\partial\ln\tau}{\partial\varepsilon}|_{\varepsilon=\varepsilon_F}=\frac{AT\mu_H\lambda}{\varepsilon_F},
\label{Eq.5}
\end{equation}
where the Hall mobility $\mu_H$=$|R_{H}|/\rho_{xx}$=$q\tau/m^*$ ($m^*$ is the effective mass of carriers). In the non-degenerate semiconducting regime, $\varepsilon_F$=$(\pi^2/3)k_BT$, and Eq.~\ref{Eq.5} is further simplified to\cite{Sun-FeSi}
\begin{equation}
\nu_N(T)=\frac{k_B}{q}\mu_H\lambda.
\label{Eq.6}
\end{equation}
Fig.~\ref{Fig.4} shows the temperature dependence of the dimensionless ratio $\lambda$ calculated from Eq.~\ref{Eq.6}. For $T$$>$10 K, $\lambda(T)$ is nearly temperature independent, with a value $\lambda_b$=$-$1.7(3), albeit the Hall mobility $\mu_H$ is strongly temperature dependent. This is consistent with the opening of a non-degenerate Kondo hybridization gap within which the Fermi energy resides. The derived value of $\lambda_b$ apparently is inconsistent with typical charge scattering processes, such as scattering by acoustic phonons with $\tau$$\propto$$\varepsilon^{-1/2}$, longitudinal optical phonons with $\tau$$\propto$$\varepsilon^{1/2}$, and by ionized impurities with $\tau$$\propto$$\varepsilon^{3/2}$\cite{Delves-Thermomag}. Instead, the enhanced magnitude of $\lambda_b$ could be due to a dispersive $\tau(\varepsilon)$ resulting from Kondo scattering\cite{Sun-CeCu2Si2}, which also leads to the large $\nu_N$. $\lambda(T)$ starts to deviate from $\lambda_b$ below 10 K, forming a sharp negative peak at 4 K where $\nu_N(T)$ also peaks with a large value 2.28 $\mu$V/(K$\cdot$T) as seen in Fig.~\ref{Fig.2}(b). This peak of $\nu_N(T)$ on top of a plateau is probably due to the coexistence of surface and bulk conductances, analogous to a multi-band picture\cite{Bel-NbSe2Nernst}. Interestingly, in the zero-temperature limit where surface conduction dominates, $\lambda(T)$ is proportional to $T$ (Fig.~\ref{Fig.4}). This deviation from a non-degenerate semiconducting regime suggests that the thermoelectric transport should be described in terms of a {\it metallic} state at low temperature (Eq.~\ref{Eq.5}). Additional evidence comes from the aforementioned linear temperature dependence of $S_{xx}(T)$ as $T$$\rightarrow$0.

In the absence of magnetic field, Eq.~\ref{Eq.2a} simplifies to
\begin{equation}
S_{xx}(T)=ATD_{xx}=AT\beta_s/\varepsilon_F,
\label{Eq.7}
\end{equation}
in which $\beta_s$ is defined as $\sigma_{xx}(\varepsilon)$$\propto$$\varepsilon^{\beta_s}$. Note that $\sigma_{xx}(\varepsilon)$$\propto$$N(\varepsilon)v^2\tau(\varepsilon)$, with $N(\varepsilon)$ being the density of states. For Dirac-fermion dispersion, $N(\varepsilon)$$\propto$$\varepsilon$, $v$=$v_F$, and $\tau$$\propto$$\varepsilon^{\lambda_s}$. Therefore we have
\begin{equation}
\beta_s=1+\lambda_s.
\label{Eq.8}
\end{equation}
Combining Eqs.~\ref{Eq.5}, \ref{Eq.7} and \ref{Eq.8}, we find $\varepsilon_F$=0.46(2) meV, $\beta_s$=0.039(2), and $\lambda_s$=$-$0.96(3). We notice that a recent ARPES measurement revealed a linear energy dependence of the quasiparticle scattering rate ($\tau^{-1}$$\propto$$\varepsilon$)\cite{Xu-SmB6ARPES2}, which is very close to our $\lambda_s$=$-$0.96(3). Additional physical parameters are then calculated and summarized in the second column of Table~\ref{Tab.1}. We emphasize here that the value of $n_s$ given in Table~\ref{Tab.1} is deduced from a fit to data shown in Fig.~\ref{Fig.2}(a) that were obtained in a Hall-bar geometry. As pointed out\cite{Wolgast-Corbino}, measurements in this geometry provide an upper bound on the real surface carrier density, and consequently, parameters derived from $n_s$ have an associated uncertainty. For example, values of $k_F$ and $m^*$ in Table~\ref{Tab.1} likely are overestimated and $v_F$ is underestimated. Taking the value for $n_s$ in Table~\ref{Tab.1} at face value, the calculated effective mass ($m^*$=$\frac{\hbar k_F}{v_F}$) for the surface state is $m^*$$\sim$17400 $m_0$ ($m_0$ is the mass of a free electron). This $m^*$ is certainly overestimated due to the uncertainty in $k_F$ as just discussed. However, considering the low $\mu_H$ and small $\varepsilon_F$, qualitatively it might be reasonable to draw the conclusion of a heavy surface state. Hall measurements on the (011)-plane of SmB$_6$ in a Corbino disc geometry, which should minimize uncertainties inherent to a Hall-bar configuration, find a surface carrier density of 2.5$\times10^{13}$ cm$^{-2}$\cite{Wolgast-Corbino} that implies a value for $k_F$ of $\sim$ 0.18 \AA$^{-1}$. This value is close to $k_F$$\sim$0.1 \AA$^{-1}$ estimated from ARPES\cite{Xu-SmB6ARPES,Hasan-SmB6ARPES,Xu-SmB6ARPES2} at the $\overline{\Gamma}$-point of the surface Brillouin zone. Taking $k_F$=0.1 \AA$^{-1}$, we estimate an effective mass of 165 $m_0$ (the third column of Table~\ref{Tab.1}). This value of $m^*$ is comparable with that of a heavy fermion metal\cite{Misra-HF} and of the theoretical estimate $m^*/m_0$$\sim$$W/\Delta_{b}$, where $W$ is the width of conduction band\cite{Alexandrov-PRL2013}. Interestingly, though the effective mass of this surface state is much larger than that of weakly correlated Bi-based TIs, the estimated metallicity parameter $k_Fl$ is comparable\cite{Ong-Bi2Te3,Ren-Bi2Te2Se,Ong-Bi2Te2Se1}.


A 2D heavy-fermion state has been proposed theoretically to exist due to strong correlations in a TKI\cite{Nikolic-HFTKI,Alexandrov-PRL2013}, but so far as we know, no direct evidence has been presented experimentally. A TKI can be regarded as a cousin of a heavy-fermion metal when SOC plays a crucial role\cite{Feng-TKI}. The Kondo hybridization between $f$- and $d$-bands gives rise to weakly dispersing renormalized narrow bands near the Fermi level. A TKI appears if the spin-orbit entangled hybridization results in a bulk gap at the Fermi level, as schematically shown in the inset to Fig.~\ref{Fig.4}. This hybridization gap is narrow, on the order of 10 meV, which has been deduced by a variety of measurements\cite{Xu-SmB6ARPES,Hasan-SmB6ARPES,Xu-SmB6ARPES2,Zhang-SmB6PCS,Ruan-SmB6STM}. In this situation, the surface states with Dirac dispersion, if they survive inside the bulk gap, will span a large Dirac cone. This is supported by the large $k_F$ measured by ARPES\cite{Xu-SmB6ARPES,Hasan-SmB6ARPES,Xu-SmB6ARPES2} as well as the small $\varepsilon_F$ determined from our thermoelectric transport measurements. Recently, Feng {\it et al.}\cite{Feng-DHF} pointed out that linear dispersion of the Dirac fermion survives inside the bulk gap only within a very small energy window, but it deforms in a larger momentum space away from the Dirac points, resulting in a large effective mass (see also in inset to Fig.~\ref{Fig.4}). These low-energy quasiparticles are therefore called {\it Dirac heavy fermions} since the contribution from $f$-electrons is included intrinsically\cite{Feng-DHF}. For comparison, we also list in the fourth column of Table \ref{Tab.1} quantities calculated in the case of 2D quadratic dispersion. The calculated effective mass is $m^*$=83 $m_0$, implying a heavy surface state in SmB$_6$. Because the nature of the surface state of a TKI is still an open question, further investigations, both theoretically and experimentally, will be beneficial.

\section{Conclusion}

To summarize, our electrical and thermoelectric transport measurements confirm the existence of a metallic surface state on the (011)-plane of SmB$_6$. The metallic surface and insulating bulk conductances are well distinguished in the thermoelectric transport. Our results demonstrate the important role played by Kondo scattering in both bulk and surface conductances and also qualitatively imply the heavy effective mass of the surface state, indicating a new intrinsic feature of interaction-driven TKIs due to the strong correlation effect in $f$-electron systems. This work also paves a way for future theoretical and experimental research.

\section*{Acknowledgments}

We are grateful to Zengwei Zhu, N. Wakeham, M. Neupane, and Z. Fisk for helpful discussions. Work at Los Alamos was performed under the auspices of the U.S. Department of Energy, Division of Materials Science and Engineering. Y. Luo acknowledges a Director's Postdoctoral Fellowship supported through the Los Alamos LDRD program. J. Dai is supported in part by NSFC (No. 11474082). Work at ZJU is supported by NSFC (No. 11190023).

\end{document}